\documentclass[%
 reprint,
 amsmath,amssymb,
 aps,
]{revtex4-1}

\usepackage{subfigure}
\usepackage{graphicx}
\usepackage{dcolumn}
\usepackage{bm}

\begin{document}

\preprint{APS/123-QED}

\title{Strong Gravitational Lensing by Static Black Holes in Effective Quantum Gravity}

\author{Yiyang Wang${}^{a, b}$}\email{numanor12138@163.com}

\author{Amnish Vachher${}^{c}$}\email{amnishvachher22@gmail.com} 
 
\author{Qiang Wu${}^{a, b}$} \email{wuq@zjut.edu.cn; Corresponding author}

\author{Tao Zhu${}^{a, b}$} \email{zhut05@zjut.edu.cn}

\author{Sushant~G.~Ghosh${}^{c, d}$}\email{sghosh2@jmi.ac.in}

 \affiliation{${}^{a}$ Institute for Theoretical Physics and Cosmology, Zhejiang University of Technology, Hangzhou 310023, China \\
${}^{b}$ United Center for Gravitational Wave Physics (UCGWP), Zhejiang University of Technology, Hangzhou, 310032, China\\
${}^{c}$Centre for Theoretical Physics, Jamia Millia Islamia, New Delhi 110025, India\\
${}^{d}$Astrophysics and Cosmology Research Unit, School of Mathematics, Statistics and Computer Science, University of KwaZulu-Natal, Private Bag 54001, Durban 4000, South Africa}

\date{\today}

\begin{abstract}
Recently, within the Hamiltonian constraint approach to Effective Quantum Gravity (EQG), two static black holes (Model-1 and Model-2 ) have been introduced that preserve general covariance. We investigate gravitational lensing in the strong-deflection regime by effective quantum gravity (EQG)-motivated black hole metrics with an additional parameter $\zeta$ besides mass $M$,  emphasizing the influence of quantum corrections on lensing phenomena. The EQG black holes encompassing the Schwarzschild hole as a particular case ($\zeta =0$) and, depending on the parameters, describe black holes with an event horizon and a Cauchy horizon (Model-1), black holes with one horizon only (Model-2) or spacetime with no horizons. Our analysis examines critical parameters such as the deflection angle, photon sphere radius, and Einstein's ring structure. Using SMBHs Sgr A* and M87* as lenses and integrating theoretical predictions with recent EHT data, we identify significant differences in lensing signatures due to quantum corrections. For Model-1, the deviations of the lensing observables: $|\delta\theta_{\infty}|$ of black holes in EQG from Schwarzschild black hole, for SMBHs Sgr A* and M87*, can reach as much as $1.75~\mu$as and $1.32~\mu$as,  while $|\delta s|$ is about $30.12$~nas for Sgr A* and $22.63$~nas for M87*. The flux ratio of the first image to all subsequent packed images indicates that EQG black hole images are brighter than their Schwarzschild counterparts, with a deviation in the brightness ratio $|\delta r_{\text{mag}}|$ reaching up to 2.02. The time delays between the second and first images, denoted $|\delta T_{2,1}|$, exhibit substantial deviations from the GR counterpart, reaching up to 1.53 \;{\rm min} for Sgr A* and 1159.9 \;{\rm min} for M87*. The EHT constraints on $\theta_{sh}$ of Sgr A* and M87* within the $1\sigma$ region limit the parameters $\zeta$. Our analysis concludes that EQG black holes are consistent with the EHT observations within this finite space. Our results show that while quantum affects the lensing observables in Model-1, Model-2 in EQG is not distinguishable from a Schwarzschild black hole in GR.
\end{abstract}

\maketitle


\section{Introduction}

Gravitational lensing, a notable prediction of Einstein's General Relativity (GR), has evolved into one of the most effective tools for studying astrophysical phenomena and testing gravitational theories. Since Einstein first considered the bending of light around massive objects \cite{Einstein:1936llh}, the field has developed significantly, leading to groundbreaking findings in both cosmology \cite{Refsdal:1964yk} and astrophysics \cite{Liebes:1964zz}. The deflection of light by a gravitational field, known as gravitational lensing, plays a critical role in disclosing the nature of spacetime and the distribution of mass in the universe.  

Black holes are one of the most fascinating predictions of GR, providing a natural arena for studying gravity in its strongest form. The theory of GR, developed by Einstein in 1915, has successfully described a wide range of astrophysical phenomena and has passed numerous observational tests across scales ranging from the solar system to cosmology \cite{Will:2014kxa}. Among these predictions, black holes—regions of spacetime with such severe gravitational fields that not even light can escape, have garnered considerable interest. Non-rotating black holes are represented by the Schwarzschild metric \cite{Schwarzschild:1916uq}, while rotating black holes are modelled by the Kerr solution \cite{Kerr:1963ud}. 

Despite its success, Einstein admitted that GR is incomplete at quantum scales, where it forecasts singularities at black hole centres and during the Big Bang. These issues and GR's incapability to reconcile with quantum mechanics offer a more comprehensive theory of gravity, especially in strong-field regimes. To handle this, several modified theories of gravity (MTG), including Loop Quantum Gravity (LQG) and string theory, have been presented \cite{Modesto:2008im, Ashtekar:2004eh, Rovelli:2014ssa, Polchinski:1998rq}. These theories offer quantum corrections to classical black hole solutions, particularly near the event horizon, resulting in modified geometries where quantum effects become prominent \cite{Gambini:2013ooa, Ghosh:2021clx, Afrin:2021wlj, Olmedo:2017lvt, Barrau:2014hda}. Such corrections may smooth out singularities and alter the behavior of spacetime in the vicinity of black holes.

 The Event Horizon Telescope (EHT) has recently provided us with the first-ever images of supermassive black holes, M87* \cite{EventHorizonTelescope:2019dse}  and Sgr A* \cite{ EventHorizonTelescope:2022xqj}, demonstrating the potential of direct observations to test the geometry of black holes and the validity of GR. These images reveal the shadows of black holes—corresponding to photon orbits trapped under the intense gravitational field—and have shown consistency with the Kerr black hole solution predicted by GR \cite{Gralla:2020yvo, Johnson:2019ljv}. However, deviations from Kerr geometry, such as those indicated by quantum gravity, could alter the black hole shadow's size and shape and influence the associated gravitational lensing effects \cite{Liu:2020ola, Tsupko:2022kwi, Gralla:2019drh, Zhu:2019ura, Liu:2020vkh, Liu:2021yev, Jiang:2024vgn, Jiang:2023img,Shi:2024bpm}.

Strong gravitational lensing, where light passes near a black hole, offers precise tests of spacetime geometry. The study of strong-gravitational lensing around black holes, particularly in the presence of quantum corrections, provides a unique window into probing the fundamental nature of gravity. In the classical regime, the bending of light near the photon sphere has been well studied using methods such as the strong deflection limit \cite{Bozza:2002af}, which provides analytical solutions for the deflection angle near black holes. Darwin \cite{unknown-author-no-date} first investigated the bending of light near a Schwarzschild black hole, showing the foundational framework for studying black hole lensing. Following developments by Virbhadra and Ellis \cite{Virbhadra:1999nm, Virbhadra:2002ju} expanded these results, specifying relativistic images formed by light passing close to the black hole's event horizon. Frittelli, Kling, and Newman \cite{Frittelli:1999yf} advanced analytical methods for solving the lens equation, which Bozza later enhanced with the strong-deflection limit (SDL) framework \cite{Bozza:2001xd, Bozza:2002zj, Bozza:2002af}. This technique has been applied to different spacetimes, such as Reissner-Nordström \cite{Eiroa:2002mk}  and rotating black holes \cite{Bozza:2002af}, facilitating a deeper understanding of gravitational lensing near black holes. The SDL has also been instrumental in studying black holes in modified gravity Models \cite{Bozza:2002zj, Eiroa:2003jf, Whisker:2004gq, Eiroa:2012fb, Bhadra:2003zs, Kumar:2022fqo, Kumar:2021cyl, Islam:2021dyk}. Recent research has focused on lensing in higher curvature gravity theories \cite{Kumar:2020sag, Islam:2020xmy, Narzilloev:2021jtg, Islam:2022ybr}, as well as modifications to Schwarzschild spacetimes \cite{Eiroa:2010wm, Ovgun:2019wej, Panpanich:2019mll, Bronnikov:2018nub, Molla:2024lpt}, revealing new phenomena and offering crucial insights into quantum gravity.

In modified gravity scenarios, quantum corrections to black hole spacetimes can affect lensing observables like deflection angles, photon spheres, and relativistic images. Recent studies in higher curvature gravity, LQG, and string theory \cite{Islam:2020xmy, KumarWalia:2022ddq, Yan_2022,Yang:2022btw,Liu:2020ola,Brahma:2020eos,Islam:2022wck,Jiang:2024cpe,Fu:2021fxn,Sahu:2015dea} show that quantum effects alter the geometry near the event horizon, influencing black hole lensing properties. For instance, quantum corrections can shrink the lensing ring and change the photon orbits, leading to deviations in the black hole shadow that can be distinguished from classical GR solutions \cite{Yang:2022btw, Zhao:2024elr, Molla:2024yde,Lu:2021htd}. Comparing forecasts of lensing in quantum-corrected spacetimes with high-resolution observations of EHT furnishes a way to test departures from GR and constrain quantum gravity \cite{EventHorizonTelescope:2019pgp, EventHorizonTelescope:2022xqj}. 

The Effective Quantum Gravity (EQG) represents one approach, incorporating quantum corrections into classical black hole solutions. Zhang et al. \cite{Zhang:2024khj}, within the Hamiltonian constraint approach to Effective Quantum Gravity (EQG), introduced two types of static black hole Models that preserve general covariance, demonstrating that quantum corrections can significantly alter black hole geometry, particularly near the event horizon.
Liu et al. \cite{Liu:2024soc} further studied the light rings and shadows of these static black holes in EQG, demonstrating that while quantum corrections do not directly affect light rings, they influence the size and characteristics of black hole shadows. It emphasizes the potential of quantum-corrected black holes to exhibit new observational signatures. In this work, we focus on spherically symmetric black holes in EQG \cite{Zhang:2024khj, Liu:2024soc}, investigating the impact of quantum corrections on lensing observables. 

We aim to explore how these corrections influence the gravitational lensing signature and discuss the prospects of testing these predictions with observations like those provided by the EHT observations. We analyze these effects using supermassive black holes like Sgr A* and M87* as lenses. While the EHT may detect strong deflection from EQG black holes, distinguishing them from Swarzschild black holes is challenging due to $\mathcal{O}(\mu$as) deviations. 

This paper is organized as follows: Section \ref{sec2} presents the framework of EQG, introducing two static spherically symmetric black holes and briefly discussing the horizon structure. Section \ref{sec3} outlines the formalism of strong gravitational lensing in the context of EQG black holes where detailed analysis of the quantum corrections on lensing observables, including deflection angles, shadows and the time delay between the first and second images on the same side of the source. In Section \ref{sec4}, we numerically analyze the strong lensing observables by taking the supermassive black holes Sgr A* and M87* as the lens. In Section \ref{sec5}, we discuss the observational prospects for detecting these quantum corrections and how current and future experiments, such as the EHT, can be used to constrain EQG Models. Finally, Section \ref{sec6} summarizes our findings and suggests directions for future research.

\section{\label{sec2}Black Hole metrics in EQG}

Zhang et al. \cite{Zhang:2024khj} derive stationary effective metrics in vacuum gravity while preserving covariance within the spherically symmetric sector. By retaining the theory’s kinematic variables and classical vector constraints, they introduce an arbitrary effective Hamiltonian constraint, $H_{\text{eff}}$, and a free function to fix the gauge related to the diffeomorphism constraint. Assuming conditions similar to classical theory, including a Dirac observable for black hole mass, they establish relationships between the effective Hamiltonian, the Dirac observable, and the free function. Solving the resulting equations yields two families of quantum-corrected metrics, reducing to classical forms when quantum parameters are zero. The metrics of static black hole in EQG \cite{Zhang:2024khj}, read:
\begin{eqnarray}
    ds_1^2=-A_1dt^2+\frac{dr^2}{B_1}+r^2d\varOmega ^2,\nonumber \\
    A_1=B_1=1-\frac{2M}{r}+\frac{\zeta ^2}{r^2}\left( 1-\frac{2M}{r} \right) ^2 ,\label{eq1}
\end{eqnarray}
and 
\begin{eqnarray}
    ds_2^2 =-A_2dt^2+\frac{dr^2}{B_2}+r^2d\varOmega ^2 \nonumber ,\\
    A_2=1-\frac{2M}{r},\; 
    B_2=A_2\left( 1+A_2\frac{\zeta ^2}{r^2} \right), \label{eq3}
\end{eqnarray}
with $\varOmega ^2 = d\theta^2+{\sin}^2\theta d\phi^2,$
and  $\zeta$ is the quantum parameter and $M$ denotes the ADM mass. These metrics will be referred to hereafter as Model-1 and Model-2.
The metric (\ref{eq1}) has two positive roots for all $M > 0$: $r_+=2M$ and 
\[
r_-=\zeta^2/\alpha-\alpha/3, \;\; \text{where}\;\alpha^3=3\zeta^2(\sqrt{81M^2+3\zeta^2}-9M. \]
These roots correspond to the event horizon ($r_+$) and the Cauchy horizon ($r_-$), respectively, depicted in figure \ref{horizons}. This characteristic reveals that spacetime (\ref{eq1}) exhibits a double-horizon structure, resembling a Reissner-Nordström black hole with a small charge. For the metric (\ref{eq3}), we get a degenerate horizon at $r=2M$.

Henceforth, we will perform strong gravitational lensing calculations for the general metric
\begin{eqnarray} \label{eqg}
  ds_1^2=-A(r) dt^2+\frac{dr^2}{B(r)}+r^2 d\varOmega ^2,  
\end{eqnarray}
and carry out numerical evaluations for Model-1 using $A=A_1$ and $B=B_1$, and similarly for Model-2 with $A=A_2$ and $B=B_2$.
 This section discusses strong-gravitational lensing by EQG black holes and analyses how the quantum parameter $\zeta$ affects the deflection angle and lensing observables. We begin by examining light propagation on the equatorial plane $(\theta=\frac{\pi}{2})$, given the spherical symmetry, in which the conservation of photon four-momentum along the Killing vectors of isometry yields two invariant quantities: energy, $\mathcal{E}$, defined as $$\mathcal{E}=-p_\mu\xi^\mu_{(t)},$$ and angular momentum $\mathcal{L}$, defined as $$\mathcal{L}=p_\mu\xi^\mu_{(\phi)},$$ where $\xi^\mu_{(t)}$ and $\xi^\mu_{(\phi)}$ are the Killing vectors associated with time and rotational invariance, respectively. Since photons follow null geodesics with $ds^2=0$, this null condition leads to
\begin{equation}\label{eq:4}
  A(r)B(r)\left(\frac{dr}{d\tau}\right)^2 +  \frac{A(r)}{r^2}\mathcal{L}^2=\mathcal{E}^2,
  \end{equation}
where the effective potential, in terms of impact parameter $u$, reads as
\begin{equation}
   V_{\text{eff}}(r)/\mathcal{E}^2= \frac{1}{A(r)B(r)}\left(\frac{A(r)u^2}{r^2}-1\right).\label{eq6}
\end{equation}
\begin{figure*}[htbp]
    \centering
    \subfigure{ 
    \includegraphics[width=0.48\linewidth]{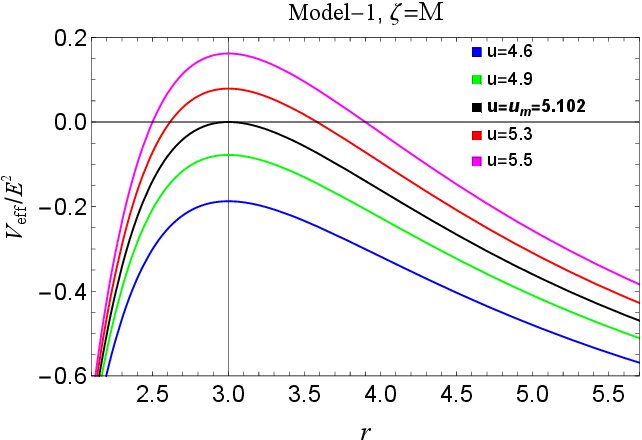}
    \label{fig 1(a)}
    }\subfigure{
    \includegraphics[width=0.48\linewidth]{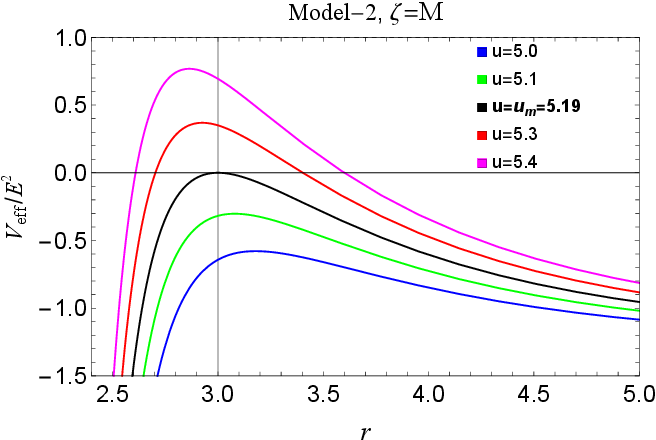}
    \label{fig 1(b)}
    }
\caption{Effective potential of photon for Model-1 and Model-2 with the quantum parameter $\zeta=M$ at marginally bound orbit. The brown curves denote the unstable circular orbit that $u_m=u\left(r_m\right)$, on which the photon can sometimes encircle around the black hole. The black hole will capture photons with $u<u_m$, while the larger ones will be scattered.}
\end{figure*}
The photon sphere exists where $V_{ \text{eff}}/\mathcal{E}^2\left( r \right)\leqslant0$.
Meanwhile, the unstable circular orbit is defined that satisfies $V_{\text{eff}}\left( r \right) =V_{\text{eff}}^{\prime}\left( r \right) =0$ and $V_{\text{eff}}^{''}\left( r \right) <0$ and it is the largest positive root called the radii of unstable photon sphere $r_m$. 
On solving, we get $r_m=3M$, which is consistent with the Schwarzschild black hole value and independent of the parameter $\zeta$.
\begin{figure*}
    \centering
    \includegraphics[width=0.48\linewidth]{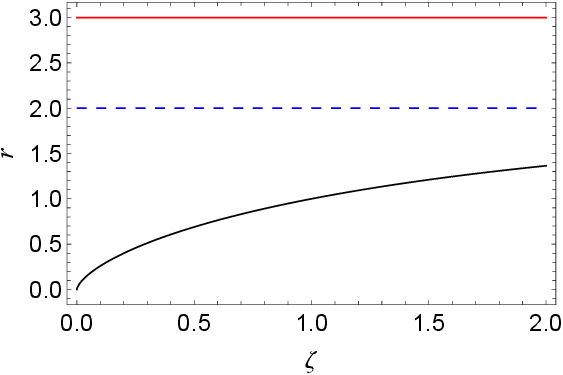}\subfigure{
    \includegraphics[width=0.5\linewidth]{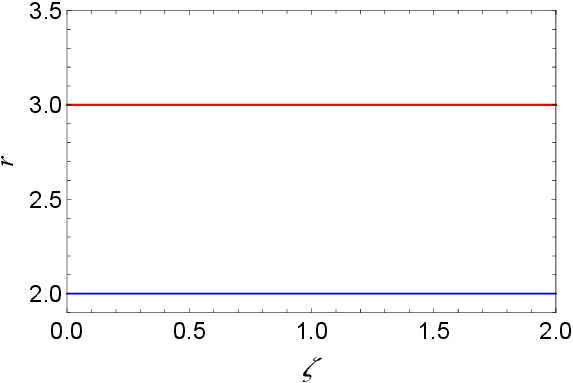}}
    \caption{Variation of horizons with parameter $\zeta$. The blue dashed line represents the Event horizon ($r_+$), while the solid black line represents the Cauchy horizon ($r_-$) for the EQG Model-1 black holes. (Left). The blue solid line represents the degenerate Event horizon ($r_+$) for EQG Model-2 black holes, and the solid red line represents the unstable photon sphere ($r_m$) in both cases (Right).}
    \label{horizons}
\end{figure*}
Throughout the trajectory that the photon moves straightforwardly to the black hole and then is redirected to the observer, the impact parameter $u$, representing the perpendicular distance from the center of the black hole to the initial direction of the photon at infinity, remains unalterable since spacetime symmetry. It can be represented as a function of the minimum approach distance $r_0$\cite{Bozza2008lens}, which vanishes the effective potential, i.e. $V_{\text{eff}}\left( r_0 \right) =0$. Thus, it can be expressed as
\begin{equation}
    u=\left| \frac{\mathcal{L}}{\mathcal{E}} \right|=\sqrt{\frac{r_0^2}{A\left( r_0 \right)}} .\label{eq8}
\end{equation}
\begin{figure}[t]
    \centering
    \includegraphics[width=\columnwidth]{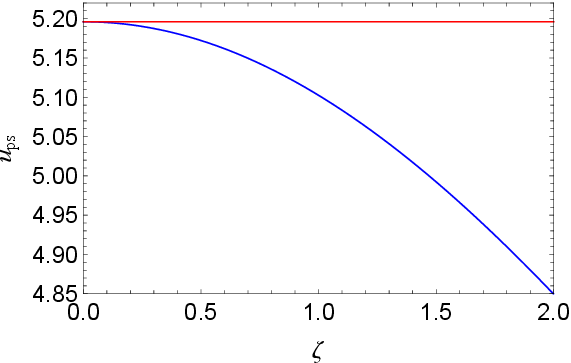}
    \caption{Variation of critical impact parameter with the parameter $\zeta$ for EQG black holes. Note that the critical impact parameters for Model-1 and Model-2 ( $\approx$ Schwarzschild black hole), are represented by blue and red lines, respectively. } \label{impact}
\end{figure}

In the limit $r_0=r_m$, the photon makes multiple loops around the black hole before reaching the observer, and the deflection angle becomes unboundedly large \cite{Virbhadra:1999nm}, the impact parameter is called the critical impact parameter($u_m$). The variation of critical impact parameter with $\zeta$ is shown in Fig.~\ref{impact}.

\section{\label{sec3}Strong Gravitational Lensing by EQG Black Holes}

The deflection angle implies significant information about gravitational lensing, and we can extract it from the photon on the equatorial plane by solving the null geodesic equation that
\begin{equation}
   \frac{d\phi}{dr}=\sqrt{\frac{Ar_{0}^{2}}{Br^2\left( A_0r^2-Ar_{0}^{2} \right)}},
\label{eq9}
\end{equation}
where the functions with subscript 0 are evaluated at $r_0$. Then the deflection angle reads as 
\begin{align}
    I\left( r_0 \right) &=2\int_{x_0}^{\infty}{\frac{d\phi}{dr}dr} ,\label{eq10}
    \\
    \alpha _D\left( r_0 \right) &=I\left( r_0 \right) -\pi. \label{eq11}
\end{align}
For all spherically symmetric metrics, the deflection angle should be logarithmic \cite{Bozza2002}, which can be rewritten in terms of impact parameter $u$ as  
\begin{equation}
    \alpha _D\left( u \right) =-\bar{a}\log \left( \frac{u}{u_m}-1 \right) +\bar{b}+\mathcal{O} \left( u-u_m \right) ,\label{eq12}
\end{equation}
with $\bar{a}$ and $\bar{b}$ denoting the lensing coefficients. To obtain these two coefficients, we define z as $z=\frac{A-A_0}{1-A_0}$, then the integral equals to
\begin{equation}
     I\left( r_0 \right) =\int_0^1{R\left( z,r_0 \right) F\left( z,r_0 \right) dz}, \label{eq13}
\end{equation}
where
\begin{align}
   R\left( z,r_0 \right) &=\frac{2\left( 1-A_0 \right) \sqrt{Ar_{0}^{2}}}{r^2A^{\prime}\sqrt{B}}, \label{eq14}
   \\
   F\left( z,r_0 \right) &=\frac{r}{r_0\sqrt{A_0-\left[ \left( 1-A_0 \right) z+A_0 \right]}}. \label{eq15}
\end{align}
The function $R\left( z,r_0 \right)$ is regular for all z, but $F\left( z,r_0 \right)$ will diverge as $z\rightarrow0$. To avoid the divergence, we expand $F\left( z,r_0 \right)$ at $z=0$ as the second Taylor series
\begin{equation}
    \bar{F}\left( z,r_0 \right) =\frac{1}{\sqrt{c_0z+d_0z^2}}. \label{eq16}
\end{equation}
To solve Eq. (\ref{eq13}), it will be split into two parts: $I_D$ acts as a divergent part, and $I_R$ is regular by subtracting the divergence from the original integral. In other word, $I\left( r_0 \right) =I_D\left( r_0 \right) +I_R\left( r_0 \right) $ with
\begin{gather}
    I_D\left( r_0 \right) =\int_0^1{R\left( 0,r_m \right) \bar{F}\left( z,r_0 \right) dz}, \label{eq17}
    \\
    I_R\left( r_0 \right) =\int_0^1{[R\left( 0,r_0 \right) F\left( z,r_0 \right) -R\left( 0,r_m \right) \bar{F}\left( z,r_0 \right)] dz} .\label{eq18}
\end{gather}
The lensing coefficients $\bar{a}$ and $\bar{b}$ are the result of the above two integrals, which can be described as
\begin{gather}
    \bar{a}=\frac{R\left( 0,r_m \right)}{2\sqrt{d_m}}, \label{eq19}
    \\
    \bar{b}=-\pi +I_R\left( r_m \right) +\bar{a}\log \left( \frac{2d_m}{A_m} \right), \label{eq20}
\end{gather}
with
\begin{equation}
   d_m=d_0|_{r_0=r_m}=\frac{r_{m}^{2}\left( 1-A_m \right) ^2\left( 2A_m-A_{m}^{''}r_{m}^{2} \right)}{4A_{m}^{2}r_{m}^{2}}. \label{eq21}
\end{equation}
   \begin{table}[b!]
   \caption{Estimates for the strong lensing coefficients $\bar{a}$, $\bar{b}$ and the critical impact parameter $u_m/R_s$ for the EQG black. Here, the value $\zeta=0$ is the Schwarzschild black hole equivalent to the EQG Model-2 black hole. Note that, $u_m$ is in the units of Schwarzschild radius $R_s = 2G M/c^2$.}
    \label{tab 1}
    \centering
    \begin{ruledtabular}
    \begin{tabular}{cccc}
          {$\zeta$} & $\bar{a}$& $\bar{b}$& $u_m/R_s$\\
         \hline

        0.0& 1.& -0.400237& 5.19615 
        \\
        0.5M& 0.981818& -0.422501& 5.17226 
        \\
        1.0M& 0.931034& -0.486982& 5.10252
        \\
        1.5M& 0.857143& -0.587184& 4.9923
        \\
        2.0M& 0.771429& -0.713767& 4.84934
        \\
    \end{tabular}
    \end{ruledtabular}
\end{table}

\begin{figure*}
   \centering
    \subfigure{ 
    \includegraphics[width=0.48\linewidth]{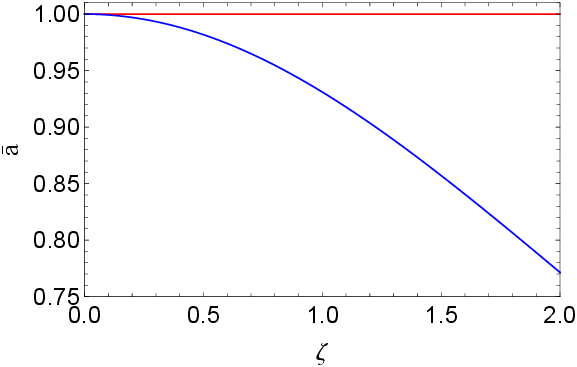}
    \label{fig4a}
    }\subfigure{
    \includegraphics[width=0.5\linewidth]{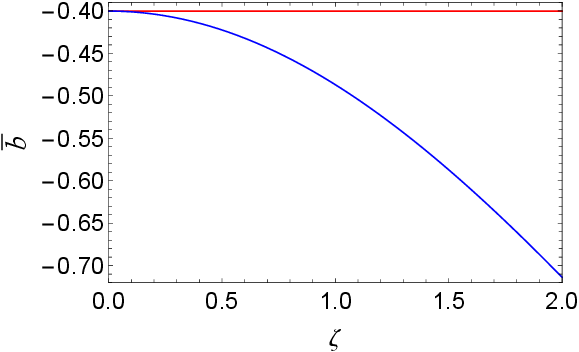}
    \label{fig4b}
    }
    \caption{Behaviour of Strong Lensing Coefficients $\bar{a}$ and $\bar{b}$ along with the parameter $\zeta$. Note that the lensing coefficients for Model-1 and Model-2 are represented by blue and red lines, respectively.} 
    \label{coeff} 
\end{figure*}
\begin{figure*}[htbp]
\centering
\subfigure{ 
\includegraphics[width=0.5\linewidth]{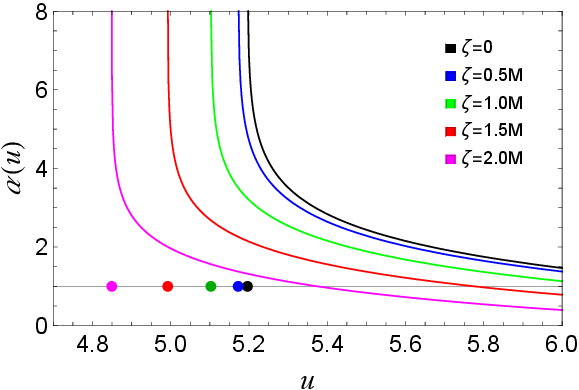}
\label{fig 2(a)}
}\subfigure{
\includegraphics[width=0.48\linewidth]{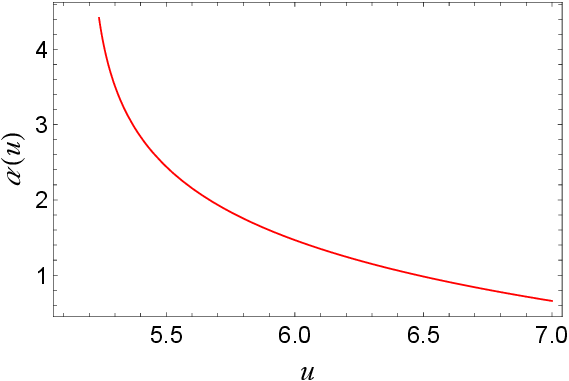}
\label{fig 2(b)}
}
\caption{(Left)Deflection angle as the function of the impact parameter $u$ for different quantum parameters $\zeta$ for EQG Model-1 black holes. The points on the horizontal axis denote the values of the impact parameter $u = u_m$ at which the deflection angle diverges. The black curves denote the Schwarzschild black hole. (Right) Deflection angle as a function of parameter $\zeta$ for EQG Model-2 black holes.}
    \label{fig 2}
\end{figure*}
\begin{figure*}[htbp]
    \centering
    \subfigure{ 
    \includegraphics[width=0.5\linewidth]{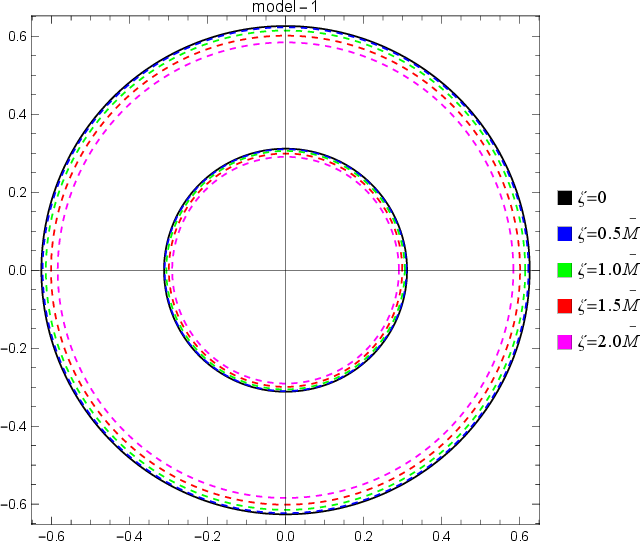}
    \label{fig 3(a)}
    }\subfigure{
    \includegraphics[width=0.43\linewidth]{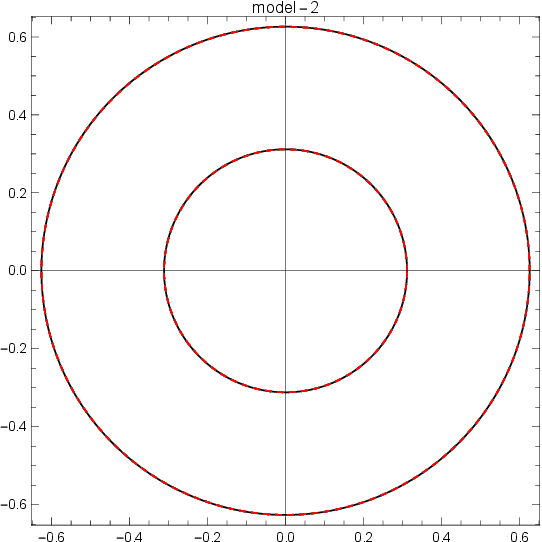}
    \label{fig 3(b)}
    }
    \caption{The outermost relativistic Einstein ring about Sgr A* (outer, $D_{\text{OL}}=9.3\; {\rm kpc}$) and M87* (inner, $D_{\text{OL}}=16.68\; {\rm Mpc}$) for Model-1 and Model-2 with different quantum parameters $\zeta$. Black curves denote the Schwarzschild black hole. For Model-2, the radius is independent of quantum parameter $\zeta$ that all rings overlap into one.}
    \label{fig 3}
\end{figure*}
We use numerical solution methods to study the relationship between the strong gravitational lensing coefficients and the quantum parameter. As seen in Fig.~\ref{coeff}, it is evident that both the deflection coefficients $\bar{a}$ and $\bar{b}$ gradually decrease as the quantum parameter $\zeta$ increases for Model-1. It is important to note that when the quantum parameter (${\zeta}=0$), the EQG BHs for Model-1 and Model-2 become equal, which also is equivalent to a Schwarzschild BH, and our calculations yield the values for a Schwarzschild black hole \cite{Bozza2002}: $\bar{a}=1$ and $\bar{b}=-0.40023$ (see Table \ref{tab 1}). Fig. \ref{fig 2} shows the typical deflection angles. With $u_0$ approaching $u_m$, the deflection angle logarithmically increases and exceeds $2\pi$, indicating that the photon closed to critical stable circular orbit sufficiently encircles the black hole as many times as it can before it's redirected to observers. As the photon is away from the $u_m$, the deflection angle turns negative, which shows that the photon is reflected opposite the black hole. But Eq. (\ref{eq12}) applies to strong gravitational lensing for the vicinity of the critical impact parameter; the deflection angle at the larger impact parameter is biased and needs to be obtained by the weak limit method. 

\begin{figure*}
\begin{center}
\includegraphics[scale=0.8]{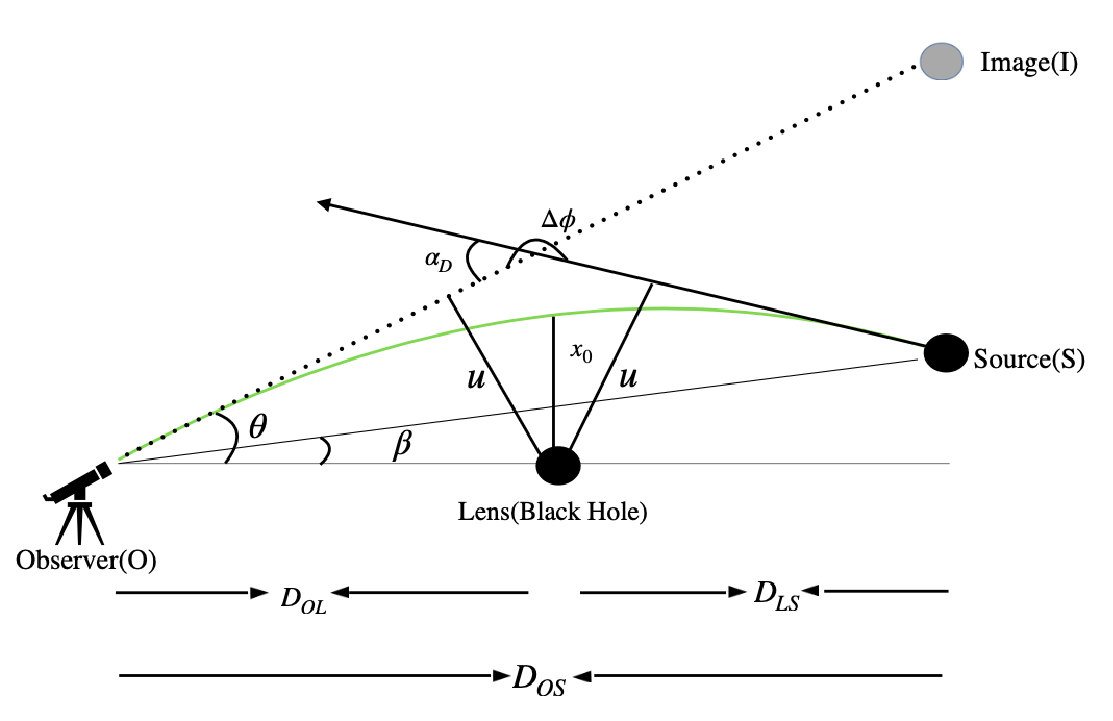}
\end{center}
\caption{  A schematic diagram illustrating the configuration of gravitational lensing, where the observer, lens, light source, and image are labeled as $\text{O}$, $\text{L}$, $\text{S}$, and $\text{I}$, respectively.\\} \label{plot5b}  
\end{figure*}

\subsection{Lens equation and observables}
Assuming that the observer and the light source are both far from the black hole, the asymptotically flat lens equation is derived as \cite{Bozza:2001xd}
\begin{equation}
    \psi =\theta -\frac{D_{\text{LS}}}{D_{\text{OS}}}\varDelta \alpha _n ,\label{eq22}
\end{equation}
where the subscripts $S$, $L$, and $O$ refer to source, lens and observer, respectively, $D$ denotes the distance and $\theta$, and $\psi$ are the angular positions of the image and source. The $\varDelta \alpha _n=\alpha _D\left( \theta \right) -2n\pi$ is the offset of deflection angle since $u=\theta D_{\text{OL}}$, where n counts the circles. Plugging the Eq. (\ref{eq12}) into Eq. (\ref{eq21}), we get \cite{Bozza:2002zj}
\begin{equation}
    \theta _n=\theta _{n}^{0}+\frac{u_me_n\left( \psi -\theta _{n}^{0} \right) D_{\text{OS}}}{\bar{a}D_{\text{LS}}D_{\text{OL}}} ,\label{eq23}
\end{equation}
where
\begin{align}
    e_n&=\exp \left( \frac{\bar{b}-2n\pi}{\bar{a}} \right), \label{eq24}
    \\
    \theta _{n}^{0}&=\frac{u_m\left( 1+e_n \right)}{D_{\text{OL}}}. \label{eq25}
\end{align}
Thus, the $n$th relativistic image magnification is \cite{Bozza:2002zj}
\begin{equation}
    \mu _n=\left.\left( \frac{\psi}{\theta}\frac{d\psi}{d\theta} \right) ^{-1}\right|_{\theta _{n}^{0}}^{}=\frac{u_{m}^{2}e_n\left( 1+e_n \right) D_{\text{OS}}}{\bar{a}\psi D_{\text{LS}}D_{\text{OL}}^{2}} ,\label{eq26}
\end{equation}
which decreases exponentially with $n$ and is inversely proportional to ${D_{OL_{}^{}}^{2}}$. Considering that the black hole, source, and observer line up and the black hole stands at the center, i.e. $\psi=0$ and $D_{\text{OS}}=D_{\text{LS}}=2D_{\text{OL}}$, then we get the Einstein ring radius yielding
\begin{equation}
    \theta _{n}^{E}=\left( 1-\frac{u_me_n}{\bar{a}D_{\text{OL}}} \right) \theta _{n}^{0} .\label{eq27}
\end{equation}
According to \cite{Chen_2019} and EHT \cite{Akiyama_2019}, we estimate the outermost Einstein ring, i.e. $n=1$, for Sgr A$^*$ with $D_{\text{OL}}=7.97 \;{\rm kpc}$ and M87* with $D_{\text{OL}}=16.8 \; {\rm Mpc}$. As $\zeta$ increases, the radius of rings both become smaller (c.f. Fig.~\ref{fig 3}). For higher order, the Einstein ring shrinks slightly.

Apart from Einstein's ring, there are still three other essential observables that $\theta _{\infty}$ represent the asymptotic position approached by a set of images that describes the separation between 
$\theta _{\infty}$ and $\theta_1$ and $r_{\text{mag}}$ is the ratio between the flux of the first image and the flux reaching from all the other images, which read as \cite{Sahu15}
\begin{align}
    \theta _{\infty}&=\frac{u_m}{D_{\text{OL}}} ,\label{eq28}
    \\
    s=\theta _1-\theta _{\infty}&\approx \theta _{\infty}\exp \left( \frac{\bar{b}-2\pi}{\bar{a}} \right) ,\label{eq29}
    \\
    r_{\text{mag}}=\frac{\mu _1}{\sum\nolimits_{n=2}^{\infty}{\mu _n}}&\approx \frac{5\pi}{\bar{a}\, \text{log}(10)}.\label{eq30}
\end{align}
\begin{figure*}
   \begin{centering}
		\begin{tabular}{c c}
		    \includegraphics[scale=0.75]{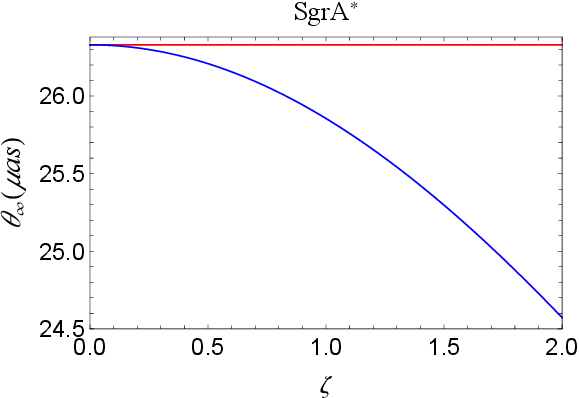}&
            \includegraphics[scale=0.75]{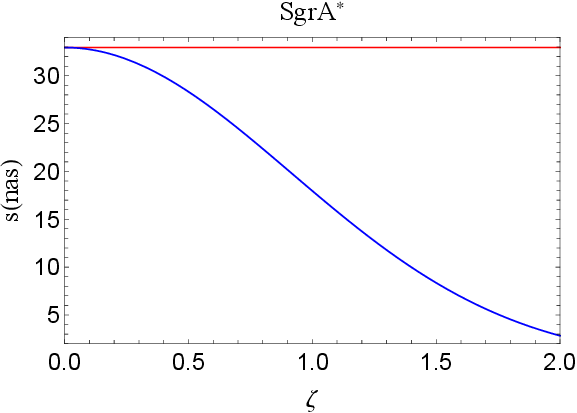}\\
			\includegraphics[scale=0.75]{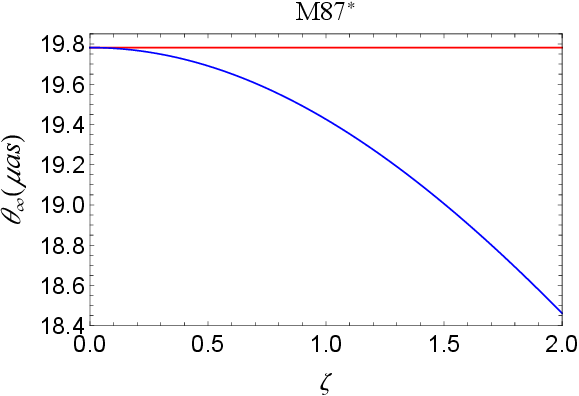}&
			\includegraphics[scale=0.75]{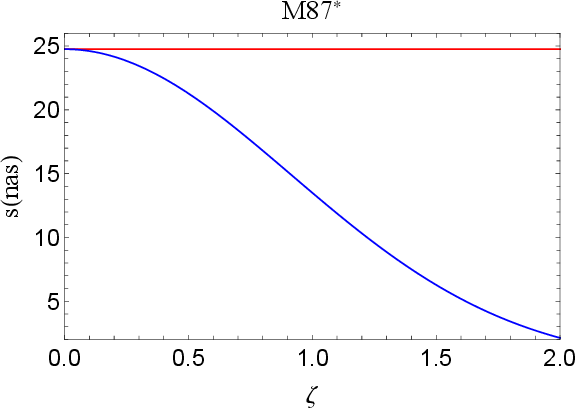}\\
			\end{tabular}
	\end{centering}
	\caption{Behaviour of Strong Lensing Observables $\theta_{\infty}$ and $s$ with the parameter $\zeta$ by taking supermassive black holes Sgr A* (Upper panel) and M87* (Lower panel) as the static black hole effective quantum gravity. Note that the lensing observables for Model-1 and Model-2 are represented by blue and red line, respectively.} \label{plottheta}
    \label{observables} 
\end{figure*}
From their astronomical observations, we can easily reconstruct the lensing parameters $\bar{a}$ and $\bar{b}$ and then examine the equation.
 \begin{table}[h!]
    \caption{The estimation of lensing observables for Sgr A* ($M=4.3\times10^6 \; {\rm M_{\odot}}$, $D_{\text{OL}}=8.3\;{\rm kpc}$) and M87* ($M=6.15\times10^9 \; {\rm M_{\odot}}$, $D_{\text{OL}}=16.68\;{\rm Mpc}$)  considered as EQG black holes for Model-1. Here, the value $\zeta=0$ corresponds to the Schwarzschild black hole, equivalent to the EQG Model-2 black hole.}
    \label{tab 2}
    \centering
    \begin{ruledtabular}
    \begin{tabular}{cccccccc}
        & \multicolumn{2}{c}{Sgr A*}& \multicolumn{2}{c}{M87*}&
        \\
        $\zeta$& $\theta_{\infty}\; ({\rm \mu as})$& $s \;({\rm nas})$& $\theta_{\infty} \;({\rm \mu as})$
        & $s \;({\rm nas})$& $r_{\text{mag}}$
        \\
        \hline 
        0.0& 26.3299& 32.9515& 19.782& 24.757& 6.82188
        \\
        0.5M& 26.2088& 28.3318& 19.6911& 21.2861& 6.94821
        \\
        1.0M& 25.8554& 17.9684& 19.4256& 13.4999& 7.32721
        \\
        1.5M& 25.2969& 8.35623& 19.006& 6.27816& 7.95886
        \\
        2.0M& 24.5725& 2.82713& 18.4617& 2.12406& 8.84318
        \\
    \end{tabular}
     \end{ruledtabular}
\end{table}

\subsection{Time delay}
Since it takes different times for the photon to encircle the black hole and then travel to various images, the time delay occurs among images. Firstly, the total time the photon spends to move from the source to the observer reads \cite{Bozza:2003cp,Virbhadra:2007kw}
\begin{equation}
    \mathrm{T}\left( \mathrm{r}_0 \right) =\tilde{\mathrm{T}}\left( \mathrm{r}_0 \right) -\int_{\mathrm{D}_{\mathrm{OL}}}^{\infty}{\left| \frac{\mathrm{dt}}{\mathrm{dr}} \right|\mathrm{dr}}-\int_{\mathrm{D}_{\mathrm{LS}}}^{\infty}{\left| \frac{\mathrm{dt}}{\mathrm{dr}} \right|\mathrm{dr}}, \label{eq31}
\end{equation}
where
\begin{equation}
   \tilde{\mathrm{T}}\left( \mathrm{r}_0 \right) =\int_{r_0}^{\infty}{\frac{2rr_0\sqrt{A_0}}{A\sqrt{A_0r^2-Ar_{0}^{2}}}dr}. \label{eq32}
\end{equation}
The integral $\tilde{\mathrm{T}}\left( \mathrm{r}_0 \right)$ describes the time photon encircling the black hole. Following the assumption that the source and observer are far from the black hole, the last two integrals cancel each other. Similar to deflection angle, $\tilde{\mathrm{T}}\left( \mathrm{r}_0 \right)$ should act logarithmically in the form of 
\begin{equation}
   \tilde{\mathrm{T}}\left( u \right) =-\tilde{a}\log \left( \frac{u}{u_m}-1 \right) +\tilde{b}+\mathcal{O} \left( u-u_m \right) ,\label{eq33} 
\end{equation}
where $\tilde{a}$ and $\tilde{b}$ are strong deflection limit coefficients \cite{Bozza:2003cp}. 

Assuming the source is variable and  we can distinguish the difference between the signal from the first and second images, then \cite{Bozza:2003cp} 
\begin{equation}
    \Delta \mathrm{T}_{2,1}^{\mathrm{s}}=2\mathrm{\pi u}_{\mathrm{m}}=2\mathrm{\pi\theta}_{\infty}\mathrm{D}_{\mathrm{\text{OL}}}.
    \label{eq34}
\end{equation}
By measuring the time delay, we can estimate the distance of the black hole.

\begingroup
\begin{table*}[tbh!]
	\caption{ Estimation and comparison of time delay for supermassive black holes at the center of nearby galaxies between the EQG Model-1 ($\zeta=2M$), EQG Model-2, and Schwarzschild black hole ($\zeta=0$). The mass ($M$) and the distance ($D_{\text{OL}}$) are given in units of solar mass and Mpc, respectively. Time delays are expressed in minutes. 
	}\label{table3} 
	\begin{ruledtabular}
		\begin{tabular}{c c c c c c c}  
				Galaxy   &           $M( M_{\odot})$      &          $D_{\text{OL}}$ (Mpc)   &     $M/D_{\text{OL}}$ & $\Delta T^s_{2,1}(\text{GR})$&$\Delta T^s_{2,1}(\text{Model-1})$ &$\Delta T^s_{2,1}(\text{Model-2})$          \\
			\hline		
			Milky Way& $  4.3\times 10^6	 $ & $0.0083 $ &       $2.471\times 10^{-11}$ & $11.4968  $ &  $10.7294 $& $11.4968  $     \\	
             M87*&$ 6.5\times 10^{9} $&$ 16.68 $
&$1.758\times 10^{-11}$& $17378.8$ &  $16218.91$& $17378.8$\\			
		
			 NGC 4472 &$ 2.54\times 10^{9} $&$ 16.72 $
&$7.246\times 10^{-12}$& $6791.11$ &  $6337.85$& $6791.11$\\
			
			 NGC 1332 &$ 1.47\times 10^{9} $&$22.66  $
&$3.094\times 10^{-12}$& $3930.29$ &  $3667.97$& $3930.29$\\
		
			 NGC 4374 &$ 9.25\times 10^{8} $&$ 18.51 $
&$2.383\times 10^{-12}$& $2473.14$ &  $2308.08$& $2473.14$\\
			
			NGC 1399&$ 8.81\times 10^{8} $&$ 20.85 $
&$2.015\times 10^{-12}$& $2355.5$ &  $2198.29$& $2355.5$\\
			 
			  NGC 3379 &$ 4.16\times 10^{8} $&$10.70$
&$1.854\times 10^{-12}$& $1112.25$ &  $1038.01$& $1112.25$\\
			
			 NGC 4486B &$ 6\times 10^{8} $&$ 16.26 $
&$1.760\times 10^{-12}$ & $ 1604.2$ &  $ 1497.13 $& $ 1604.2$\\
		
			 NGC 1374 &$ 5.90\times 10^{8} $&$ 19.57 $ &$1.438\times 10^{-12}$& $1577.46$ &  $1472.18$& $1577.46$\\
			    
			NGC 4649&$ 4.72\times 10^{9} $&$ 16.46 $
&$1.367\times 10^{-12}$& $ 12619.7$ &  $ 11777.4 $& $ 12619.7$\\
		
			NGC 3608 &$  4.65\times 10^{8}  $&$ 22.75  $ &$9.750\times 10^{-13}$& $1243.26$ &  $1160.28$& $1243.26$\\
		
			 NGC 3377 &$ 1.78\times 10^{8} $&$ 10.99$
&$7.726\times 10^{-13}$ & $475.913$ &  $444.149$& $475.913$\\
		
			NGC 4697 &$  2.02\times 10^{8}  $&$ 12.54  $ &$7.684\times 10^{-13}$& $540.081$ &  $504.034$& $540.081$\\
			 
			 NGC 5128 &$  5.69\times 10^{7}  $& $3.62   $ &$7.498\times 10^{-13}$& $152.132$ &  $141.978$& $152.132$\\
			
			NGC 1316&$  1.69\times 10^{8}  $&$20.95   $ &$3.848\times 10^{-13}$& $451.85 $ &  $421.69$& $451.85 $\\
			
			 NGC 3607 &$ 1.37\times 10^{8} $&$ 22.65  $ &$2.885\times 10^{-13}$& $366.292 $ &  $341.845$& $366.292 $\\
			
			NGC 4473 &$  0.90\times 10^{8}  $&$ 15.25  $ &$2.815\times 10^{-13}$& $240.63$ &  $224.57$& $240.63$\\
			
			 NGC 4459 &$ 6.96\times 10^{7} $&$ 16.01  $ &$2.073\times 10^{-13}$ & $186.087 $ &  $173.667$& $186.087 $\\
		
			M32 &$ 2.45\times 10^6$ &$ 0.8057 $
&$1.450\times 10^{-13}$ & $6.55048 $ &  $6.11328 $  & $6.55048 $  \\
			
			 NGC 4486A &$ 1.44\times 10^{7} $&$ 18.36  $ &$3.741\times 10^{-14}$ & $38.5008$ &  $35.93$& $38.5008$\\
			 
			NGC 4382 &$  1.30\times 10^{7}  $&$ 17.88 $  &$3.468\times 10^{-14}$& $34.7577 $ &  $32.4378$& $34.7577 $\\
		
			CYGNUS A &$  2.66\times 10^{9}  $&$ 242.7 $  &$1.4174\times 10^{-15}$& $7111.95 $ &  $6637.28$& $7111.95 $\\
		\end{tabular}
	\end{ruledtabular}
\end{table*}
\endgroup

\section{\label{sec4}Lensing effects of\\ the supermassive black holes \\Sgr A* and M87*}

In this section, we estimate and compare the lensing observables $\theta_{\infty}$, $s$, and $r_{\text{mag}}$ for EQG black holes with those of Schwarzschild black holes, using the supermassive black holes M87* and Sgr A* as candidates for EQG black holes. Using the latest astronomical observation data, the mass and distance from Earth for M87* are approximately $\left(6.5\pm0.7\right)\times{10}^9 \,{\rm M_{\odot}}$ and $16.8$ \,{\rm Mpc} \cite{EventHorizonTelescope:2019pgp,EventHorizonTelescope:2019ggy} respectively. For Sgr A* they are about $4.0_{-0.6}^{+1.1}\times{10}^6 \; {\rm M_{\odot}}$ and $7.97$ \; {\rm kpc} respectively. \cite{EventHorizonTelescope:2022wkp,EventHorizonTelescope:2022apq}.

We calculated the angular positions of the first- and second-order relativistic primary and secondary images for Sgr A* and M87* using Eq.~(\ref{eq23}), assuming $d = D_{\text{LS}}/D_{\text{OS}} = 1/2$. Our results show that in EQG black holes, the angular positions of the images are smaller than in the Swarzschild black hole and exhibit minimal sensitivity to the source position $\psi$. For larger $\psi$ values, $\theta_{1p} > |\theta_{1s}|$, while for smaller $\psi$, the values converge closely. This behavior persists across higher-order relativistic images. In EQG, the first and second-order primary images are approximately 1.78 \;{\rm$\mu$as}  and 1.75 \;{\rm$\mu$as} larger than their GR counterparts at $\zeta = 2M$. However, this difference is too small to be detected with current telescopes, especially given the strong demagnification of these relativistic images.

Comparing the strong lensing observations for Model-2 of EQG black holes with those for Model-1, which shows similar behavior as in the Schwarzschild black hole case ($\zeta=0$), we observe that the angular separation $s$ between images is smaller. At the same time, their relative magnification $r_{\text{mag}}$ is larger compared to the Schwarzschild case. For Sgr A*, the angular position $\theta_\infty$ is in the range of (24.57, 26.33) \;{\rm$\mu$as}, while for M87* it is in the range of (18.46, 19.78) \;{\rm$\mu$as}. The angular separation $s$ for SMBHs Sgr A* ranges from (2.8327-32.95) nas and for M87* from (2.12-24.75) nas. The deviations of the lensing observables $|\delta s|$ and $|\delta \theta_\infty|$ for EQG black holes from the Schwarzschild black hole can reach up to 30.12 nas and 1.75 \;{\rm$\mu$as} for Sgr A*, and 22.63 nas and 1.32 \;{\rm$\mu$as} for M87*, respectively. Furthermore, the angular separation between relativistic images rendered by the Sgr A* black hole exceeds that of the M87* black hole for fixed parameter values. Therefore, EQG images outshine Schwarzschild's black hole images, a phenomenon apparent from the increasing flux ratio of the first image relative to all other images as $\zeta$ increases. Using Eq.~(\ref{eq26}), the relative magnification of the first and second-order images are tabulated in Table \ref{table6} and \ref{table7} for black holes in GR and EQG at $(\zeta=2M)$. The first-order images in EQG are highly magnified compared to the second-order images and the corresponding images in GR. The brightness of EQG black hole images compared to Schwarzschild black hole increases with higher values of $\zeta$, as shown in Fig.~\ref{observables}. In Table \ref{table3}, we have gathered updated data on 14 supermassive black holes with varying masses and distances from Sgr A*. Using this data, we calculated the time delays between the first and second-order relativistic primary images ($\Delta T^s_{2,1}$). For Sgr A* and M87*, the time delay can reach $\sim  10.73\;{\rm min}$ and $\sim 16218.9\;{\rm min}$ at $\zeta=2M$, respectively. These values deviate from their corresponding black holes in General Relativity (GR) by $\sim 0.7674\; {\rm min}$ for Sgr A* and  $\sim 1159.9\;{\rm min}$  for M87*. Although these deviations are insignificant for Sgr A*, they become significant for M87* and other black holes, offering valuable insights for testing and comparing EQG black holes with those predicted by general relativity.
\begin{table}[h]
    \caption{The deviation of Schwarzschild black hole from the EQG black holes of Model-1 for different values of parameter $\zeta$ has been calculated where $\delta\theta_{\infty}=\theta_{\infty}(EQG)-\theta_{\infty}(Sch)$ and $\delta s ({\rm nas})=s_{\text{EQG}}-s_{\text{Sch}}$. Note that for EQG Model-2 $\delta\theta_{\infty}$ and $\delta s ({\rm nas})$ is $0$}
    \label{tab3}
    \centering
    \begin{ruledtabular}
    \begin{tabular}{cccccccc}
        & \multicolumn{2}{c}{Sgr A*}& \multicolumn{2}{c}{M87*}&
        \\
        $\zeta$& $\delta\theta_{\infty} ({\rm \mu as})$& $\delta s ({\rm nas})$& $\delta\theta_{\infty} ({\rm \mu as})$
        & $\delta s ({\rm nas})$& $\delta r_{\text{mag}}$
        \\
        \hline 
        0.5M& -0.1211& -4.6197& -0.091& -3.4709& 0.1263
        \\
        1.0M& -0.4745& -14.9831& -0.3564& -11.2571& 0.5053
        \\
        1.5M& -1.033& -24.5953& -0.776& -18.4789& 1.137
        \\
        2.0M& -1.7574& -30.1244& -1.3203& -22.633& 2.0213
        \\
    \end{tabular}
    \end{ruledtabular}
\end{table}

\begingroup
\begin{table*}
	\caption{
		{\bf Image positions of first and second order primary and secondary images due to lensing by Sgr A* with $d=D_{\text{LS}}/D_{\text{OS}}=1/2$}: EQG Model-2 which is equivalent to Schwarzschild black hole and EQG Model-1 ($\zeta=2M)$ predictions for angular positions $\theta$ of primary ($p$)   and secondary images ($s$)  are given for different values of angular source position $\psi$. {\bf (a)} All angles are in $\mu${\em as}. {\bf (b)} We have used $M_{\text{Sgr A*}}= 4.3\times 10^6	\, {\rm m}$, $D_{\text{OL}}= 8.3 \times 10^{6} \, {\rm pc}$.
	}\label{table4}
	\begin{ruledtabular}
		\begin{tabular}{l cccc cccc}
			&
			\multicolumn{4}{c}{ EQG}&
			\multicolumn{4}{c}{GR}\\
$\psi$&$\theta_{1p,{\rm EQG}} $&$\theta_{2p,{\rm EQG}}$&$\theta_{1s,{\rm EQG}}$&$\theta_{2s,{\rm EQG}}$&$\theta_{1p,{\rm GR}} $&$\theta_{2p,{\rm GR}}$&$\theta_{1s,{\rm GR}}$&$\theta_{2s,{\rm GR}}$\\
			\hline
		
            $0$& 24.5753 & 24.5725 & -24.5753 & -24.5725 & 26.3628 & 26.3299 & -26.3628 & -26.3299 \\
			$10^{0} $& 24.5753 & 24.5725 & -24.5753 & -24.5725 & 26.3628 & 26.3299 & -26.3628 & -26.3299 \\
			$10^{1} $ & 24.5753 & 24.5725 & -24.5753 & -24.5725 & 26.3628 & 26.3299 & -26.3628 & -26.3299 \\
			$10^{2} $ & 24.5753 & 24.5725 & -24.5753 & -24.5725 & 26.3628 & 26.3299 & -26.3628 & -26.3299 \\
			$10^{3} $ & 24.5754 &24.5725 & -24.5753 & -24.5725 & 26.3631 & 26.3299 & -26.3625 & -26.3299 \\
			$10^{4} $  & 24.5757 & 24.5725 & -24.575 & -24.5725 & 26.366 & 26.3299 & -26.3596 & -26.3299 \\
		\end{tabular}
	\end{ruledtabular}
\end{table*}
\endgroup

\begingroup
\begin{table*}
	\caption{
		{\bf Image positions of first and second order primary and secondary images due to lensing by M87* with $d=D_{\text{LS}}/D_{\text{OS}}=1/2$}: EQG Model-2 which is equivalent to Schwarzschild black hole and EQG Model-1 ($\zeta=2M)$ predictions for angular positions $\theta$ of primary ($p$)   and secondary images ($s$)  are given for different values of angular source position $\psi$. {\bf (a)} All angles are in \, {\rm $\mu$as}. {\bf (b)} We have used $M_{\text{M87*}}=  6.5 \times 10^9 \, {\rm m}$, $D_{\text{OL}}=  16.8  \times 10^6 \, {\rm pc}$.
	}\label{table5}
	\begin{ruledtabular}
		\begin{tabular}{l cccc cccc}
			&
			\multicolumn{4}{c}{ EQG}&
			\multicolumn{4}{c}{GR}\\
$\psi$&$\theta_{1p,{\rm EQG}} $&$\theta_{2p,{\rm EQG}}$&$\theta_{1s,{\rm EQG}}$&$\theta_{2s,{\rm EQG}}$&$\theta_{1p,{\rm GR}} $&$\theta_{2p,{\rm GR}}$&$\theta_{1s,{\rm GR}}$&$\theta_{2s,{\rm GR}}$\\
			\hline
	
            $0$& 18.4638 & 18.4617 & -18.4638 & -18.4617 & 19.8068 & 19.7821 & -19.8068 & -19.7821 \\
			$10^{0} $&  18.4638 & 18.4617 & -18.4638 & -18.4617 & 19.8068 & 19.7821 & -19.8068 & -19.7821 \\
			$10^{1} $ & 18.4638 & 18.4617 & -18.4638 & -18.4617 & 19.8068 & 19.7821 & -19.8068 & -19.7821 \\
			$10^{2} $ & 18.4638 & 18.4617 & -18.4638 & -18.4617 & 19.8068 & 19.7821 & -19.8068 & -19.7821 \\
			$10^{3} $ &  18.4639 & 18.4617 & -18.4638 & -18.4617 & 19.807 & 19.7821 & -19.8065 & -19.7821 \\
			$10^{4} $  &  18.4641 & 18.4617 & -18.4636 & -18.4617 & 19.8092 & 19.7821 & -19.8044 & -19.7821 \\
		\end{tabular}
	\end{ruledtabular}
\end{table*}
\endgroup

\begin{table*}[t]
	\caption{
		{\bf Magnifications of first order and second order relativistic images due to lensing by Sgr A* with $d=D_{\text{LS}}/D_{\text{OS}}=1/2$}: EQG Model-2 which is equivalent to Schwarzschild black hole and EQG Model-1 ($\zeta=2M)$ predictions for magnifications $\mu_n$ is given for different values of angular source position $\psi$. {\bf (a)} $1p$ and $1s$ refer to first order relativistic images on the same side as primary and secondary images, respectively. {\bf (b)}  We have used $M_{\text{Sgr A*}}= 4.3\times 10^6	\, {\rm m}$, $D_{\text{OL}}= 8.3 \times 10^{6} \, {\rm pc}$ .
	}\label{table6}
\resizebox{\textwidth}{!}{
 \begin{centering}	
	\begin{tabular}{p{0.8cm} p{2.5cm} p{2.5cm} p{2.7cm} p{2.7cm} p{2.5cm} p{2.2cm} p{2.7cm} p{2.2cm}}
\hline\hline
	&
			\multicolumn{4}{c}{EQG}&
			\multicolumn{4}{c}{GR}\\
    $\psi$&  $\mu_{1p,{\rm EQG}} $ & $ \mu_{2p,{\rm EQG}}$&$\mu_{1s,{\rm EQG}}$&$\mu_{2s,{\rm EQG}}$&$\mu_{1p,{\rm GR}}$&$\mu_{2p,{\rm GR}}$&$\mu_{1s,{\rm GR}}$&$\mu_{2s,{\rm GR}}$\\
	\hline 	\hline

		
	    	$10^{0}  $&$8.73281 \times 10^{-13}$&$ 2.53415  \times	10^{-16} $&$-8.73281 \times 10^{-13} $&$ -2.53415  \times	10^{-16} $&$8.52495 \times 10^{-12}  $&$1.59  \times	10^{-14} $&$-8.52495 \times 10^{-12}  $&$-1.59  \times	10^{-14}  $\\
			
			$10^{1}  $&$8.73281 \times 10^{-14}$&$ 2.53415  \times	10^{-17} $&$-8.73281 \times 10^{-14} $&$ -2.53415  \times	10^{-17} $&$8.52495 \times 10^{-13}$&$1.59  \times	10^{-15} $&$-8.52495 \times 10^{-13}  $&$- 1.59  \times	10^{-15}	$\\
			
			$10^{2}  $&$ 8.73281 \times 10^{-15} $&$ 2.53415 \times	10^{-18} $&$-8.73281 \times 10^{-15} $&$ -2.53415  \times	10^{-18} $&$8.52495 \times 10^{-14} $&$1.59  \times	10^{-16} $&$-8.52495 \times 10^{-14} $&$ -1.59  \times	10^{-16}	$\\
			
			$10^{3}  $&$8.73281 \times 10^{-16} $&$ 2.53415  \times	10^{-19} $&$-8.73281 \times 10^{-16}$&$ -2.53415  \times	10^{-19} $&$8.52495 \times 10^{-15}  $&$ 1.59  \times	10^{-17} $&$-8.52495 \times 10^{-15}  $&$ -1.59  \times	10^{-17} 		$\\
			
			$10^{4}  $&$8.73281 \times 10^{-17} $&$  2.53415  \times	10^{-20} $&$-8.73281 \times 10^{-17} $&$ -2.53415  \times	10^{-20} $&$8.52495 \times 10^{-16} $&$1.59  \times	10^{-18}$&$-8.52495 \times 10^{-16}  $&$ -1.59  \times	10^{-18} 	$\\
\hline\hline
	\end{tabular}
\end{centering}
}	
\end{table*}

\begin{table*}[t]
	\caption{
		{\bf Magnifications of first and second order relativistic images due to lensing by M87* with $d=D_{\text{LS}}/D_{\text{OS}}=1/2$}: EQG Model-2 which is equivalent to Schwarzschild black hole and EQG Model-1 ($\zeta=2M)$ predictions for magnifications $\mu_n$ are given for different values of angular source position $\psi$. {\bf (a)} $1p$ and $1s$ refer to first order relativistic images on the same side as primary and secondary images, respectively. {\bf (b)}  We have used $M_{\text{M87*}}=  6.5 \times 10^9 \, {\rm m}$, $D_{\text{OL}}=  16.8  \times 10^6 \, {\rm pc}$. 
	}\label{table7}
\resizebox{\textwidth}{!}{
 \begin{centering}	
	\begin{tabular}{p{0.8cm} p{2.5cm} p{2.5cm} p{2.7cm} p{2.7cm} p{2.5cm} p{2.5cm} p{2.7cm} p{2.5cm}}
\hline\hline
	&
			\multicolumn{4}{c}{EQG}&
			\multicolumn{4}{c}{GR}\\
    $\psi$&  $\mu_{1p,{\rm EQG}} $ & $ \mu_{2p,{\rm EQG}}$&$\mu_{1s,{\rm EQG}}$&$\mu_{2s,{\rm EQG}}$&$\mu_{1p,{\rm GR}}$&$\mu_{2p,{\rm GR}}$&$\mu_{1s,{\rm GR}}$&$\mu_{2s,{\rm GR}}$\\
	\hline 	\hline

		
			$10^{0}$  & $4.92945 \times 10^{-13}$ & $1.43046)  \times	10^{-16}$ &$-4.92945  \times 10^{-13} $&$ -1.43046  \times	10^{-16} $&$4.75466  \times 10^{-12}  $&$8.86797 \times	10^{-15} $&$-4.75466 \times	10^{-12}  $&$-8.86797  \times	10^{-15}  $\\
			
			$10^{1}  $&$4.92945\times10^{-14}$&$1.43046  \times	10^{-17} $&$ -4.92945\times10^{-14}$&$ -1.43046  \times	10^{-17} $&$4.75466\times10^{-13}$&$8.86797 \times	10^{-16} $&$-4.75466\times10^{-13}  $&$ -8.86797  \times	10^{-16}	$\\
			
			$10^{2}  $&$4.92945\times10^{-15} $&$1.43046  \times	10^{-18} $&$ -4.92945\times10^{-15}$&$ -1.43046  \times	10^{-18} $&$ 4.75466\times10^{-14} $&$8.86797  \times	10^{-17} $&$-4.75466\times10^{-14} $&$ -8.86797  \times	10^{-17}	$\\
			
			$10^{3}  $&$4.92945\times10^{-16} $&$ 1.43046  \times	10^{-19} $&$-4.92945\times10^{-16}$&$-1.43046  \times	10^{-19} $&$ 4.75466\times10^{-15}  $&$ 8.86797 \times	10^{-18} $&$-4.75466 \times 10^{-15}  $&$ -8.86797  \times	10^{-18}		$\\
			
			$10^{4}  $&$4.92945\times10^{-17} $&$1.43046 \times	10^{-20} $&$-4.92945\times10^{-17} $&$ -1.43046  \times	10^{-20} $&$4.75466\times10^{-16} $&$8.86797 \times	10^{-19} $&$-4.75466\times10^{-16}  $&$ -8.86797 \times	10^{-19}  		$\\
\hline\hline
	\end{tabular}
\end{centering}
}	
\end{table*}

\section{\label{sec5}EHT Constraints on the Observables}

In 2019, the Event Horizon Telescope (EHT) Collaboration made a groundbreaking achievement by capturing the horizon-scale image of the supermassive black hole M87* \citep{EventHorizonTelescope:2019dse,EventHorizonTelescope:2019pgp,EventHorizonTelescope:2019ggy}. With a distance of $d=16.8\;  {\rm Mpc}$ and an estimated mass of $M=(6.5 \pm 0.7) \times 10^9  \; {\rm M_{\odot}}$, the EHT Collaboration determined that M87* has a compact emission region with an angular diameter of $\theta_d=42\pm 3$\;{\rm $\mu$as}. The central flux depression, which indicates the shadow of the black hole, is at least a factor of 10.
In 2022, the EHT observed the supermassive black hole Sgr A* at the center of our Milky Way, revealing an image with a diameter ring $\theta_d= 48.7 \pm 7\,\mu$as. This observation suggests a black hole mass of $M = 4.0^{+1.1}_{-0.6} \times 10^6  \; {\rm M_{\odot}} $ and a Schwarzschild shadow deviation $\delta = -0.08^{+0.09}_{-0.09}~\text{(VLTI)},-0.04^{+0.09}_{-0.10}~\text{(Keck)}$. These findings provide a valuable new method to probe strong field gravity \citep{EventHorizonTelescope:2022wkp,EventHorizonTelescope:2022urf,EventHorizonTelescope:2022xqj}. The EHT also calculated an angular diameter of the emission ring for Sgr A * of $\theta_d=(51.8\pm 2.3)$\;{\rm $\mu$as} based on the previously estimated mass $M$ and $D_{\text{LS}}=8.15\pm 0.15$ \; {\rm kpc} \citep{EventHorizonTelescope:2022xqj}.

The horizon-scale images of these supermassive black holes offer a novel theoretical approach to testing gravity theories \cite{Kumar:2020yem,Ghosh:2020syx,Kumar:2018ple,Afrin:2021imp,EventHorizonTelescope:2021dqv,EventHorizonTelescope:2022xqj,Afrin:2021wlj,Ghosh:2022kit,Islam:2022ybr,KumarWalia:2022aop,Zakharov:2021gbg,Vagnozzi:2022moj,Saurabh:2020zqg,Kumar:2020hgm,Vachher:2024fxs,Afrin:2021wlj,Vagnozzi:2022moj,Afrin:2021imp,Vachher:2024ldc}. By analyzing the EHT observations of the shadows of Sgr A* and M87*, we can constrain the deviation parameter $\zeta$ for the EQG black hole spacetime of Model-1. Using the apparent radius of the photon sphere $\theta_\infty$, as a proxy for the angular size of the black hole shadow, we can place constraints on this deviation parameter within the 1-$\sigma$ confidence level.   
\begin{figure*}[tbh!]
	\begin{centering}
		\begin{tabular}{c c}
\includegraphics[scale=0.95]{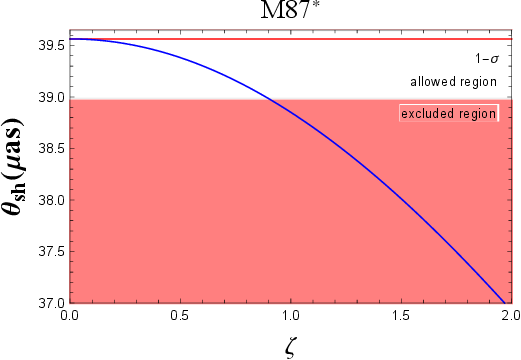}\hspace{0.5cm}
\includegraphics[scale=0.95]{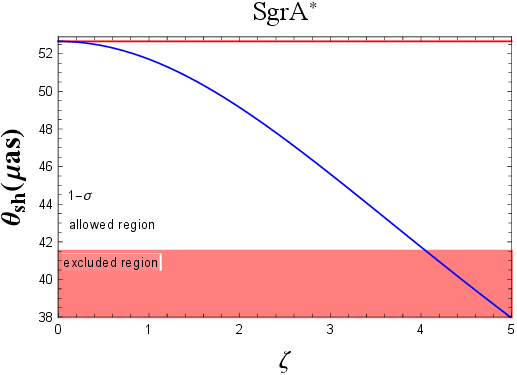}
			\end{tabular}
	\end{centering}
\caption{Shadow angular diameter $\theta_{\text{sh}}(=2\theta_{\infty})$ as a function of $\psi$, when M87* and Sgr A* is Modeled as EQG black holes . The blue and red lines correspond to calculated values of $2\theta_{\infty}$ as a function of $\zeta$ for Model-1 and Model-2, respectively.  The white/red regions represent areas 1 $\sigma$ consistent/inconsistent with the EHT observations, indicating that the latter imposed constraints on parameter $\zeta$.}\label{SgrAparameter}
\end{figure*}
\paragraph{Constraints from  M87*:} We find that the Schwarzschild black hole ($\zeta = 0$) casts the largest shadow with its angular diameter $\theta_{\text{sh}}=2\theta_\infty= 39.56069$\;{\rm $\mu$as}, which falls within the 1 $\sigma$ region for the black hole with mass  $M=(6.5 \pm 0.7) \times 10^9  \; {\rm M_{\odot}}$ and distance of $D_{\text{OL}}=16.8$ \; {\rm Mpc} \citep{EventHorizonTelescope:2019dse,EventHorizonTelescope:2019pgp,EventHorizonTelescope:2019ggy}. Figure \ref{SgrAparameter} depicts the angular diameter $\theta_{\text{sh}}$ as a function of $\zeta$, with the black solid line corresponding to $\theta_{\text{sh}}=39~\mu$as for the EQG black hole spacetime as M87*. The EQG black hole spacetime metric, when investigated with the EHT results of M87*  within the 1 $\sigma$ bound, constrains the parameter $\zeta$, viz., $0< \zeta \le 0.9 M$. Thus, based on Fig.~\ref{SgrAparameter}, EQG black hole spacetime could be a candidate for astrophysical black holes within only a tiny part of the parameter space.  

\paragraph{Constraints from  Sgr A*:}The EHT observation used three independent algorithms to find that the mean measured value of the shadow angular diameter, which lies in the range of $\theta_{\text{sh}} \in (46.9, 50)$\;{\rm $\mu$as}, and the 1 $\sigma$ interval is $\in$ $(41.7,55.6)$\;{\rm $\mu$as} \citep{EventHorizonTelescope:2022xqj}. The angular diameter $\theta_{\text{sh}} \in (41.7,55.6)$\;{\rm $\mu$as}, which falls within the 1 $\sigma$ confidence region with the observed angular diameter of the EHT observation of the Sgr A* black hole, constrains the parameter $0.0 \le \zeta \le 4 M$ for the EQG black hole spacetime.  Thus, within the finite parameter space, the EQG black hole spacetime agrees with the EHT results of the Sgr A* black hole shadow (see Fig.~\ref{SgrAparameter}).

\section{\label{sec6}Conculsion}

Recent developments in the EQG framework have introduced significant modifications to black hole spacetimes through quantum corrections. Zhang et al. \cite{Zhang:2024khj} derived stationary effective metrics while preserving covariance in spherical symmetry. By retaining the kinematic variables and classical vector constraints, they formulated an arbitrary effective Hamiltonian constraint $H_{\text{eff}}$ and a free function addressing the gauge associated with the diffeomorphism constraint. It yielded two families of effective Hamiltonian constraints, each characterized by a quantum parameter. When these parameters vanish, the constraints revert to their classical forms, leading to two quantum-corrected metrics with distinct spacetime structures. Building on this, we explored strong-gravitational lensing by these black holes (Model-1 and Model-2) that generalize the Schwarzschild black hole by introducing the parameter $\zeta$. Our analysis shows that quantum corrections significantly affect the lensing observables for Model-1, while Model-2 remains indistinguishable from the Schwarzschild solution.

We examined the impact of $\zeta$ on lensing coefficients $(\bar{a}, \bar{b})$ and deflection angle. For Model-1, the coefficients decrease with increasing $\zeta$ (Fig.~\ref{fig 2}), while Model-2 shows no dependence on $\zeta$. The deflection angle $\alpha_D$ for a fixed impact parameter is smaller for EQG black holes than for GR ones.
Using supermassive black holes M87* and Sgr A* as lenses, we found that quantum corrections lead to deviations in key lensing features, such as angular positions, image brightness, and time delays. The brightness of the first image increases, while the time delays between subsequent images also increase. Our results indicate that the angular position $\theta_\infty$ and the image separation $s$  decrease as $\zeta$ increases, while the brightness ratio $r_{\text{mag}}$ increases. The Einstein ring for EQG black holes is smaller than that for the GR black hole.

We found that the relativistic image angular position $\theta_\infty$ varies between $19.782 \,{\rm \mu\text{as} }\geq \theta_\infty(\text{M87*}) \geq 18.4617 \,{\rm \mu\text{as} }$ and $26.3299 \,{\rm \mu\text{as} }\geq \theta_\infty(\text{Sgr A*}) \geq 24.5725 \,{\rm \mu\text{as} }$. The deviation $|\delta\theta_\infty|$ can reach $1.75 \,{\rm \mu\text{as} }$ for Sgr A* and $1.32 \,{\rm \mu\text{as} }$ for M87*. The time delays between the second and first images, $\Delta T_{2,1}$, show deviations from the GR counterpart, reaching up to 0.7674 \,{\rm min} for Sgr A* and 1159.9 \;{\rm min} for M87* . EHT observations place constraints on the EQG parameter $\zeta$, with bounds $0 < \zeta \leq 0.9 M$ for M87* and $0 \leq \zeta \leq 4 M$ for Sgr A*, indicating that the M87* constraint is tighter.

In conclusion, our analysis shows that gravitational lensing is a promising tool for testing quantum corrections in EQG theory. The constraints emanated from the EHT data suggest that EQG black holes align with current observations within a finite range of $\zeta$. With future high-precision observations, especially from the forthcoming ngEHT,  these constraints could be further refined, potentially differentiating EQG black holes from classical Schwarzschild solutions. It offers an exciting avenue to deepen our understanding of black hole physics and quantum gravity.

\begin{acknowledgments}
This work is supported by the National Key Research and Development Program of China Grant No. 2020YFC2201503. S.G.G. and A.V. would like to thank SERB-DST for project No. CRG/2021/005771. T.Z. is also partly supported by the Zhejiang Provincial Natural Science Foundation of China under Grant Nos. LR21A050001 and LY20A050002, the National Natural Science Foundation of China under Grant No. 12275238.
\end{acknowledgments}

%

\end{document}